\documentclass[12pt]{article}
\usepackage[yyyymmdd,hhmmss]{datetime}
\usepackage{graphicx}
\usepackage{pdfpages}
\usepackage{hyperref}
\usepackage{amssymb}
\usepackage{color}
\usepackage{amsmath}
\numberwithin{equation}{section}
\usepackage{stackengine}
\stackMath
\newcommand\tsup[2][2]{%
 \def\useanchorwidth{T}%
  \ifnum#1>1%
    \stackon[-.5pt]{\tsup[\numexpr#1-1\relax]{#2}}{\scriptscriptstyle\sim}%
  \else%
    \stackon[.5pt]{#2}{\scriptscriptstyle\sim}%
  \fi%
}

\def\text#1{\mbox{#1}}

\newcommand{\be}{\begin{equation}}
\newcommand{\ee}{\end{equation}}
\newcommand{\beq}{\begin{eqnarray}}
\newcommand{\eeq}{\end{eqnarray}}

\def\tr{{\rm Tr}}

\newcommand{\veq}{\mathrel{\rotatebox{90}{$=$}}}

\newcommand{\ldeq}{\mathrel{\rotatebox{45}{$=$}}}
\newcommand{\rdeq}{\mathrel{\rotatebox{-45}{$=$}}}

\begin{document}

\begin{center}

{\bf \Large
Path integral of free fields
and the determinant of Laplacian
in warped space-time
}

\vspace{1cm}

Soumangsu Chakraborty$^{1,2}$, Akikazu Hashimoto$^3$, Horatiu Nastase$^4$

\vspace{1cm}

{}$^1$ Universit\'e Paris-Saclay, CNRS, CEA, \\Institut de Physique Th\'eorique, 91191 Gif-sur-Yvette, France

{}$^2$ Department of Physics\\
Center for Cosmology and AstroParticle Physics (CCAPP)\\
The Ohio State University\\
191 W Woodruff Ave, Columbus, OH 43210, USA

{}$^3$ Department of Physics, University of Wisconsin \\ 1150 University Ave, Madison, WI 53706, USA

{}$^4$ Instituto de Física Teórica, UNESP-Universidade Estadual Paulista\\
R. Dr. Bento T. Ferraz 271, Bl. II, Sao Paulo 01140-070, SP, Brazil

\end{center}

{\it E-mail:} \href{mailto:soumangsuchakraborty@gmail.com}{soumangsuchakraborty@gmail.com}, 
\href{mailto:hashimoto@wisc.edu}{aki@physics.wisc.edu},
\href{mailto:horatiu.nastase@unesp.br}{horatiu.nastase@unesp.br}

\vspace{1cm}

\noindent We revisit the problem of computing the determinant of Klein-Gordon operator $\Delta = -\nabla^2 + M^2$ on Euclideanized $AdS_3$ with the Euclideanized time coordinate compactified with period $\beta$,  $H_3/Z$, by explicitly computing its eigenvalues and computing their product.  Upon assuming that eigenfunctions are normalizable on $H_3/Z$, we found that there are no such eigenfunctions. Upon closer examination, we discover that the intuition that $H_3/Z$ is like a box with normalizable eigenfunctions was false, and that there is, instead, a set of eigenfunctions which forms a continuum.  Somewhat to our surprise, we find that there is a different operator $\tilde \Delta = r^2 \Delta$, which has the property that (1) the determinant of $\Delta$ and the determinant of $r^2 \Delta$ have the same dependence on $\beta$, and that (2) the Green's function of $\Delta$ can be spectrally decomposed into eigenfunctions of $\tilde \Delta$. We identify the $\tilde \Delta$ operator as the ``weighted Laplacian'' in the context of warped compactifications, and comment on possible applications.

\newpage

\tableofcontents

\section{Introduction}

Anti-de Sitter space in 2+1 dimensions is an important geometry in the study of gravity and string theory. A quantum theory of gravity on $AdS_3$ is conjectured to be dual to a conformal field theory in two dimensions via the holographic correspondence.   String theory in some $AdS_3$ background supported by NSNS fluxes can be analyzed using traditional world sheet techniques. Global $AdS_3$ at finite temperature and the BTZ black hole geometry have the same Euclidean continuation. These are some of the well-established features of $AdS_3$ which continue to motivate further study.

In this article, we re-visit the computation of the determinant of the Klein-Gordon operator
\be \Delta = -\nabla^2 + M^2~, \ee
on compactified Euclidean $AdS_3$ which we denote as $EAdS_3/Z = H_3/Z$ where we periodically identify the Euclidean time coordinate $\tau \sim \tau + \beta$. This quantity is closely related to the one-loop vacuum amplitude of free scalar living on $H_3/Z$. It has been considered recently by  \cite{Giombi:2008vd,David:2009xg,Denef:2009kn,Castro:2017mfj,Keeler:2018lza,Kakkar:2022hub}, although  there are even earlier works \cite{Avis:1977yn,Burgess:1984ti,Camporesi:1991nw,Camporesi:1994ga,Ichinose:1994rg,Mann:1996ze}. 

In the treatment of \cite{Camporesi:1991nw,Camporesi:1994ga,Mann:1996ze,Giombi:2008vd,David:2009xg}, the heat kernel method, which we will briefly review below, was used. These quantities are expected to be related to various quantities outlined in figure \ref{figa}. One important test of the computation using the heat kernel method is that the inverse square root of the determinant is found to agree with the Boltzmann sum of the spectrum of states for free scalars on $AdS_3$ at temperature $T=1/\beta$. 
\begin{figure}[h]
\centerline{
\begin{tabular}{ccccc}
&&\fbox{\parbox{1.3in}{Heat kernel \\
$\exp\left[ {1 \over 2}  \int {dt \over t} {\bf K}(t) \right]$}} &= &\fbox{\parbox{1.7in}{Boltzmann-like sum\\ of quasi-normal modes\\ in BTZ}}\\
& $\ldeq$ & $\veq$ & $\rdeq$ & $\veq$ \\
\fbox{\parbox{1.3in}{Boltzmann sum\\ $Z(\beta) = \sum_i e^{-\beta E_i}$ \\ on $AdS_3$}} & 
$=$ & 
\fbox{\parbox{1.5in}{Path integral \\ $\int [D\Phi] e^{-{1 \over 2} \int d^dx\, \sqrt{g}\Phi \Delta \Phi}$ \\ on $H_3/Z$}} & {$=$} &
\fbox{\parbox{1.3in}{``Boltzmann sum''\\ $Z(\beta) = \sum_i e^{- E_i/\beta}$ \\ on BTZ}}   \\
$ \veq$ &   & { $\veq$} & & $\veq$ \\
 \fbox{\parbox{1.0in}{$(\det r^2 \Delta)^{-1/2}$\\ on $H_3/Z$}} & {=} & \fbox{\parbox{1.0in}{$(\det \Delta)^{-1/2}$ \\ on $H_3/Z$}} &= & \fbox{\parbox{1.5in}{$(\det (r^2-1) \Delta)^{-1/2}$\\ on $H_3/Z$}}
\end{tabular}
}
\caption{Standard relation between the Boltzmann sum, the expression based on the heat kernel, the path integral of a free boson, and the determinant of $\Delta$, as well as the expected relation to thermodynamics on BTZ.  $E_i$ is the energy of the multi-particle state for a gas of free particles. The equality with $\det (r^2 \Delta)$ is one of the surprises we report in this article. The ``Boltzmann sum'' for BTZ is in quotes because the spectrum is actually continuous and as such the sum is somewhat formal, as we discuss in appendix \ref{appb}. Similar continuum appears in the spectrum of $\Delta$ on $H_3/Z$ as we will describe in section \ref{sec:res}. \label{figa}}
\end{figure}

The heat kernel method is a powerful technique to compute determinants without actually computing the eigenvalues explicitly. (See also \cite{Dunne:2007rt} for a discussion of the Gel’fand-Yaglom theorem in similar spirits.) We decided nonetheless to try to compute the determinant by actually computing the eigenvalues, and computing their product. Part of our motivation is to relate this exercise to a controlled prescription for performing the path integral mode by mode.  In the course of performing this exercise, we encountered a puzzle. The goal of this article is to explain the puzzle and its resolution. In the course of this discussion, we also encountered a surprise. 

The puzzle, the resolution, and the surprise will be described in detail in the following section, but we can summarize its content as follows. To determine the eigenvalues of Klein-Gordon operator $\Delta$, we looked for normalizable modes $\Delta \psi = \lambda \psi$ on $H_3/Z$ with the expectation that $H_3/Z$ should behave like a ``box'' in three dimensions. The puzzle is that we do not find such normalizable modes. The resolution is that $H_3/Z$ is not like a ``box'' and that the modes on $H_3/Z$ form a continuum. (They are $\delta$-function normalizable.) The surprise is that there is an alternate operator $\tilde \Delta= r^2 \Delta$ whose spectrum is discrete and also appears to have the property that $\det \Delta = \det \tilde \Delta$. This alternate operator $\tilde \Delta$ is an object known as the ``weighted Laplacian'' in the context of studying warped compactifications \cite{DeLuca:2024fbc}. We also find that the weighted Laplacian is useful for computing the Green's function of $\Delta$.  

The paper is organized as follows. In section \ref{sec:background}, we review the heat kernel approach to compute the partition function of a free scalar field in $H_3/Z$. In section \ref{sec:puzzle},
we discuss a puzzle encountered in computing the eigenfunctions of the Klein-Gordon operator in $H_3$ with a decaying boundary condition and in section \ref{sec:res} we provide a resolution to the puzzle discussed in section \ref{sec:puzzle}. We show by explicit computation that this apparent paradox is due to a false assumption on the harmonic analysis in $H_3$. In section \ref{sec:surprise}, we introduce a new operator, the ``weighted Laplacian," that one encounters in the study of warped compactifications. We show that the new operator accepts eigenfunctions that are normalizable and has a discrete spectrum, and exactly reproduces the partition function of a free scalar field in $H_3/Z$. We also show that the Green's function associated with the standard Klein-Gordon operator in $H_3$ admits a spectral decomposition in terms of the eigenfunctions of the ``weighted Laplacian". Finally in section \ref{sec:discussion}, we discuss the implications of our findings and potential avenues for future research.

\section{Background} \label{sec:background}

In order to explain the puzzle, it is useful to first recall the standard relations between the path integral of Klein-Gordon fields, 
the determinant of Klein-Gordon operator $\Delta$, and 
an expression based on the heat kernel, 
on a generic Euclidean manifold ${\cal M}$. 

The standard explanation \cite{Coleman:1985rnk} for the relation between the path integral of the Klein-Gordon field and the determinant of $\Delta$ is as follows. One first determines the eigenfunctions
\be \Delta \Phi_\lambda(x) = \lambda \Phi_\lambda(x) \ . \ee
and parameterize the generic field configuration by the coefficient $a_\lambda$ via relation
\be \Phi(x) = \sum_\lambda a_\lambda \Phi_\lambda(x) \ . \ee
One then writes
\be \int [D \Phi(x)] e^{-{1 \over 2} \int dx \, \sqrt{g} \Phi \Delta \Phi} = \int [da_\lambda] e^{-\sum_\lambda a_\lambda \lambda^2} =  \prod_\lambda \lambda^{-1/2} = (\det \Delta)^{-1/2} \ . \label{eigenprod} \ee

The relation between the heat kernel and the determinant of $\Delta$ is as follows. The heat kernel $K(x,x',t)$ is defined by a differential equation 
\be \Delta K(x,x',t) = -{\partial \over \partial t} K(x,x',t)~, \ee
where $K$ is defined on ${\cal M} \times {\cal M} \times R$, with the boundary condition that
\be K(x,x',t=0) = {1 \over \sqrt{g}} \delta(x-x') \ . \ee
Then, a formal solution is
\be K(x,x,'t) = e^{-t \Delta} \ . \ee
Then, we have that
\be \int {dt \over t} \mbox{Tr} K(x,x,'t) = \int {dt \over t} \mbox{Tr} e^{-t \Delta} = -\mbox{Tr} \log \Delta =- \log (\det \Delta)~,\   \label{Kdet} \ee
where 
\be  \mbox{Tr} K(x, x', t) = \int dx \,  \sqrt{g}  K(x,x,t) \equiv {\bf K}(t) \ . \ee
So the path integral, the exponential of an integral the trace of the heat kernel, and the inverse square root of the determinant of $\Delta$ evaluate to the same quantity. 

Finally, if ${\cal M}$ has some isometry $d/d\tau$, one can consider periodically identified manifold ${\cal M}/Z$. Then, the inverse square root of the determinant is expected to agree with the thermal Boltzmann sum of the set of states arising from a free scalar field living in the space-time obtained by Wick rotating coordinate $\tau = i \tilde \tau$ on ${\cal M}$. 

We can summarize the equality of four quantities by a set of six qualities illustrated in figure \ref{figa}.

Let us now consider the case where ${\cal M} = H_3$ is the Euclideanized $AdS_3$. We will write the metric of $H_3$ as
\be ds_{H_3}^2 = r^2 d \tau^2 + (r^2 - 1) d \theta^2 + {dr^2 \over r^2 - 1}~ , \qquad \theta = \theta + 2 \pi \ . \label{h3metric} \ee
Coordinates $(\tau, \phi, r)$ are dimensionless, and $r \ge 1$. Coordinate $\theta = \theta+ 2 \pi$ is periodic. This will make the point at $r=1$ singularity free. 
The shift by $d/d\tau$ is an isometry. We will specify $H_3/Z$ as the (\ref{h3metric}) subjected to periodicity condition 
\be \tau \sim \tau+\beta \ . \ee
The analytic continuation $\tau = i \tilde \tau$ will lead to the usual $AdS_3$ space-time with the metric 
\be ds_{AdS_3}^2 = -r^2 d \tilde \tau^2 + (r^2 - 1) d \theta^2 + {dr^2 \over r^2 - 1}~ , \qquad \theta = \theta + 2 \pi \ . \label{ads3metric} \ee

The spectrum and the normal modes have been computed by many. See e.g. (2.19) of \cite{Fitzpatrick:2010zm}. Earlier analysis can be found e.g. in  \cite{Witten:1998zw}. There are infinitely many single-particle states with energies
\be E = \ell + \ell'  + 2h~, \qquad \ell, \ell' \ge 0 \in Z ~,\label{Ellp} \ee
and
\be h = {1 \over 2} (1 + \sqrt{1 + M^2})~. \ee
The Boltzmann sum for the gas of these states on $AdS_3$ is given by the Planck spectrum 
\be Z_{AdS_3}(\beta) = \prod_{\ell, \ell' =0}^\infty {1 \over 1 - q^{2h +\ell+\ell'}}~, \qquad q = e^{-\beta} \ . \label{ZAH3} \ee
This result can be compared against the analysis of \cite{Mann:1996ze,Giombi:2008vd} which primarily computed the heat kernel ${\bf K}(t)$ on $H_3/Z$ and found (see (4.9) of \cite{Giombi:2008vd})
\be {\bf K}(t) =  2 \sum_{n=1}^\infty {(2 \pi \beta) (2 \pi) \over 4 |\sinh \pi n \beta|^2}  {e^{-(M^2+1)t - {(2 \pi n \beta)^2 \over 4t}} \over 4 \pi ^{3/2} \sqrt{t}}~. \label{res2}\ee
Finally, with a little bit of algebra (see (4.10) of \cite{Giombi:2008vd}), it was shown that 
\be \exp\left[ {1 \over 2}  \int {dt \over t} {\bf K}(t) \right] = Z_{AdS_3}(\beta)  \ . \ee
This confirms the leftmost vertical equality in figure \ref{figa}.  If we can take all of the equalities in figure \ref{figa} for granted, then it follows that the computations (\ref{ZAH3}) and (\ref{res2}) are two independent ways of computing the determinant of $\Delta$ on $H_3/Z$ and the partition function.

\section{Puzzle \label{sec:puzzle}}

We are now ready to explain the puzzle, which arose from asking the following simple question. ``What happens if one tries to compute the determinant of $\Delta$ as the product of eigenvalues of $\Delta$?'' 

In order to determine the eigenvalues, one sets up the eigenvalue problem which is simply the differential equation
\be \Delta \Phi_\lambda(x_i) = \lambda \Phi_\lambda(x_i) \ . \label{eigen1} \ee
Using coordinates $x_i = (\tau, \theta, r)$, and the metric (\ref{h3metric}),  eigenequation (\ref{eigen1}) reads
\be \left(-{1 \over \sqrt{g}} (\partial_\tau \sqrt{g} g^{\tau\tau} \partial_\tau+\partial_\theta \sqrt{g} g^{\theta\theta} \partial_\theta+\partial_r \sqrt{g} g^{rr} \partial_r) + M^2 \right)\Phi_\lambda(\tau, \theta, r) = \lambda \Phi_\lambda(\tau,\theta,r)~, \ee
Since the metric (\ref{h3metric}) is invariant under translation in $\tau$ and $\theta$ with periodicities $2 \pi \beta$ and $2 \pi$ respectively, it is natural to parameterize
\be
\Phi(\tau,\theta,r) = e^{i m_1 \tau/\beta + i m_2 \theta} g(r)  \label{parameterize} \ . 
\ee
so that the equation for $g(r)$ becomes
\be -\left(r^2-1\right)
   g''(r)+\frac{\left(1-3
   r^2\right) g'(r)}{r}+\frac{g(r)
   \left(\frac{m_1^2}{\beta^2}+\frac{m_2^
   2 r^2}{r^2-1}\right)}{r^2}+M^2
   g(r) 
= \lambda g(r)~.  \label{waveeq3} \ee
We can set $M^2=0$ for the time being and bring it back at the end since it only shifts $\lambda$. Although (\ref{waveeq3}) looks somewhat complicated, it admits a closed-form solution in terms of hypergeometric functions. All that remains to be specified in order to enumerate the set of eigenvalues and eigenfunctions is to impose a set of boundary conditions on $g(r)$. The variable $r$ ranges from 1 to $\infty$, with $r=\infty$ corresponding to the boundary of $H_3/Z$. 

The eigenvalue problem is the problem of finding the set of allowed $\lambda$ which solves (\ref{waveeq3}) subject to the appropriate boundary condition.  One is therefore led to consider what boundary condition one should be imposing.

One can try to arrive at the requisite boundary condition as follows. Suppose we reparameterize the radial coordinate 
\be y = {1 \over r^2 - 1}~, \ee
which maps the boundary of $H_3/Z$ to $y=0$ and $r=1$ to $y=\infty$. Equation (\ref{waveeq3}) now reads
\be 
0 = g(y) \left(- \lambda + y \left(\frac{m_1^2}{\beta^2 (y+1)}+m_2^2\right)\right)-4 y^2 \left((y+1) g''(y)+g'(y)\right) \ .  \label{waveeq4}
\ee
These solutions to these equations near $y=0$ scale as
\be g(y) \sim y^{(1 \pm  \sqrt{1 -  \lambda})/2} \ . \label{smally} \ee
Since we often think of anti-de Sitter spaces as a box, it seems natural to require, assuming that the real part of $\sqrt{1 -  \lambda}$ is positive, for ``$-$'' component to vanish. This  imposes one boundary condition at $y=0$, and we find that given $\lambda$, $g(y)$ is uniquely determined up to an overall multiplicative factor
\beq
&&g(y) = y^{\frac{1}{2} \left(\sqrt{1- \lambda }+1\right)}
   (y+1)^{\frac{i m_1}{2
   \beta}} \, _2F_1\left(a,b,c,-y\right)~, \label{gy}
\eeq
with
\beq
a & = &\frac{1}{2} \left(- m_2+\sqrt{1-
   \lambda }+\frac{ \sqrt{-m_1^2
   \beta^2}}{\beta^2}+1\right)~,  \cr
   b & = &\frac{1}{2} \left(m_2+\sqrt{1-
   \lambda }+\frac{ \sqrt{-m_1^2
   \beta^2}}{\beta^2}+1\right)~,
\cr
c& = & \sqrt{1-\lambda}+1~. \eeq   
Let us now explore the behavior of (\ref{gy})  near $y=\infty$. For this, we use 
\beq _2 F_1(a,b,c,z)& = &z^{-a} \left(\frac{(-1)^{-a} \Gamma (b-a) \Gamma (c)}{\Gamma (b) \Gamma (c-a)}+O\left(\left(\frac{1}{z}\right)^1\right)\right)\cr
&& +z^{-b} \left(\frac{(-1)^{-b}
   \Gamma (a-b) \Gamma (c)}{\Gamma (a) \Gamma (c-b)}+O\left(\left(\frac{1}{z}\right)^1\right)\right) ~.\label{largey} \eeq
For our solution
\be a - b = - m_2~, \ee
and so for $m_2 > 0$, $a$ is smaller, and $z^{-a}$ term is the dominant one. We want to enforce that dominant component to vanish. This will happen for
\be b = - N~, \label{bc1} \ee
or 
\be c - a = - N~, \label{bc2}\ee
These equations read
\be \sqrt{1-
   \lambda } = -2 N - m_2 -1\mp\frac{ \sqrt{-m_1^2
   \beta^2}}{\beta^2} \ . \ee
This equation does not admit any solution, since we assumed that $\sqrt{1- \lambda }$ and $m_2$ are both non-negative.\footnote{We thank S. Giombi for discussion on this point.}  The set of eigenmodes of $\Delta$ decaying at the boundary and regular at the center that we set out to find do not appear to exist.  How can one then compute the eigenvalues and the determinant of $\Delta$?

\section{Resolution \label{sec:res}}

In this section, we describe the resolution of the puzzle outlined in section \ref{sec:puzzle}. Our strategy is to take a closer look at the harmonic equation (\ref{waveeq3}). One way to extract qualitative intuition about equations of this type is to manipulate it into the standard Schr\"{o}dinger form. There is a systematic procedure for achieving this. 

\begin{figure}
\centerline{\includegraphics[width=3in]{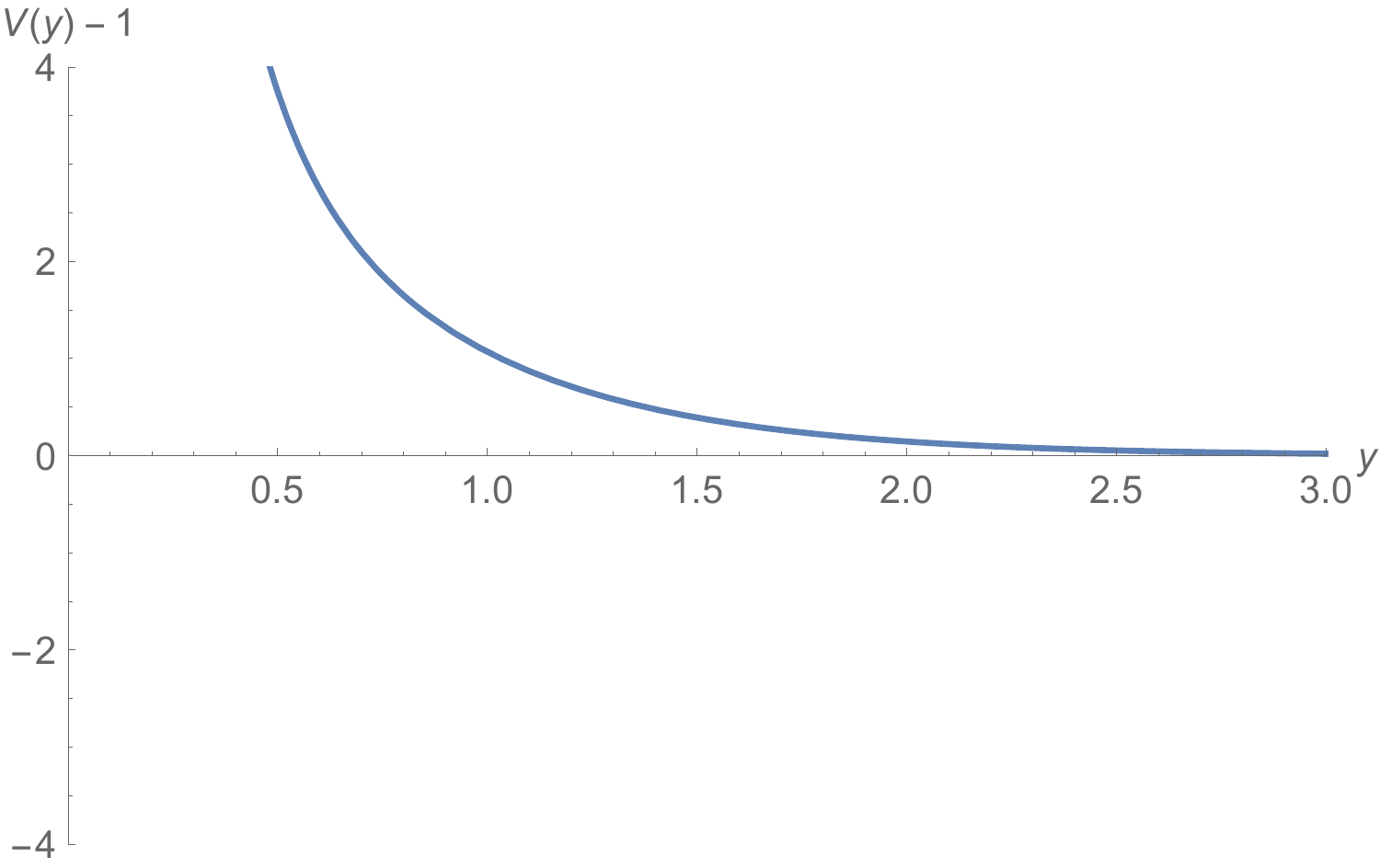}}
\caption{Potential $V(y)$ for $m_1=m_2=1$, $M=0$, and $\beta = 1$. It is clear that one dimensional Schr\"{o}dinger problem with this potential does not admit any bound state solution. \label{figd}}
\end{figure}

First, normalize the equation so that the $\lambda$ term is independent of $r$. Equation (\ref{waveeq3}) is already in that form. Next, we look for a change of variable
\be y=y(r)~, \ee
which will bring the coefficient of the $g''(y)$ term to be $-1$.  This is achieved by using the variable
\be y = \log(r + \sqrt{r^2-1})~, \ee
so that $r=1$ is mapped to $y=0$ whereas $r=\infty$ is mapped to $y=\infty$. The differential equation is then of the form
\be -g''(y) + P(y) g'(y) + Q(y)g(y) = \lambda g(y)~. \ee
Now, we can set
\be g(y) = f(y) \tilde g(y) ~,\ee
and chose $f(y)$ such that the coefficient of $\tilde g'(y)$ term vanishes. This is achieved by setting
\be f(y) = {1 \over \sqrt{2 \sinh (2y)}} ~,\ee
making the wave equation take the Schr\"{o}dinger form
\be - \tilde g''(y) + V(y) \tilde g(y) + M^2 \tilde g(y)= \lambda \tilde g(y) ~,\ee
where
\be V(y) = \left(\frac{m_1^2}{\tau_2^2}+\frac{1}{4}\right)
   \text{sech}^2(y)+\left(m_2^2-\frac{1}{4}\right) \text{csch}^2(y)+1 \label{Vy}~.
   \ee
If one sets $m_1 = m_2 = 1$, the potential $V(y)$ takes the form illustrated in figure \ref{figd}

We see from this potential that a wavefunction decaying at the boundary ($y=\infty$) and regular at the origin $(y=0)$ with real $\lambda$ can not exist. There will, however, be a continuum set of delta function normalizable solutions for $\lambda \ge 1 + M^2$. 

The fact that the spectrum on $H_3$ is continuous above the threshold $\lambda = 1+M^2$ is also consistent with the heat kernel analysis. Let us for the time being consider the non-compact case. This simply amounts to taking the same differential equation (\ref{waveeq3}) and sending $\beta=\infty$ keeping $E = m_1/\beta$ fixed. So we expect sets of eigenfunctions
\be \Delta \psi_\lambda = \lambda \psi_\lambda \ . \ee
Since the set of eigenfunctions is now continuous, some care is needed in specifying the normalization convention for $\psi_\lambda(x)$ where $x \in H_3$.  Strictly speaking, this is an arbitrary choice as long as one is careful about the fact that various intermediate formulas will take different forms in different sets of conventions. Let us choose the normalization convention so that
\be \psi_\lambda(r=1) = 1 \ . \label{normconv}\ee
This is the normalization convention adopted in \cite{Camporesi:1991nw} but not \cite{Camporesi:1994ga}. With normalization convention (\ref{normconv}) specified, the LHS of the equation
\be \int dx\, \sqrt{g} \psi_\lambda(x) \psi^*_{\lambda'}(x) = {1 \over \rho(\lambda)} \delta(\lambda - \lambda')~, \label{thenorm} \ee
is fully prescribed, from which we can read off the function $\rho(\lambda)$. This quantity is related to the spectral function discussed in equation (2.10)  of \cite{Camporesi:1994ga} and (6.9) of \cite{Kapec:2024zdj}.\footnote{In order to make the relation precise, one must account for the fact that $\lambda_{here} = \lambda_{CH}^2$. This implies  $\rho(\lambda)=\lambda^{-1/2}  \mu(\sqrt{\lambda})$. Equation (6.9) and (6.11) of \cite{Kapec:2024zdj} can be seen to be equivalent to (\ref{normconv}) and (\ref{thenorm}) when one accounts for the difference in the normalization convention of the eigenfunctions.} Once $\rho(\lambda)$ is determined, we can also write the other orthonormality relation, 
\be \int d \lambda \, \rho(\lambda)\psi_\lambda(x) \psi^*_\lambda(x') = {1 \over \sqrt{g}} \delta(x-x') \ . \ee

With the normalization convention for the eigenfunctions specified, one can show that
\be K(t,x,x') = \int d \lambda \, e^{-\lambda t} \rho(\lambda) \psi_\lambda(x) \psi^*_\lambda(x')~, \ee
is the heat kernel, since
\be \Delta K = \lambda K = -{\partial \over \partial t} K \ , \ee
and
\be K(0,x,x') = \int d \lambda \, \rho(\lambda) \psi_\lambda(x) \psi^*_\lambda(x') = {1 \over \sqrt{g}} \delta(x-x')~.\ee

It is convenient to define the resolvent
\be W(\lambda, x, x' ) = \int dt\, K(t,x,x') e^{\lambda t} = \int d\lambda' {\rho(\lambda') \over \lambda' - \lambda} \psi_{\lambda'}(x) \psi^*_{\lambda'}(x')~. \ee
Using the fact that 
\be {1 \over x} = P\left[{1 \over x} \right] - i \pi \delta(x)\ , \ee
where  $P$ stands for principle value, we have
\be \mbox{Im} W(\lambda, x, x' ) = -\pi  \rho(\lambda)  \psi_{\lambda'}(x) \psi^*_{\lambda}(x') \ .  \ee
Finally, we use the fact that $K(t,x,x')$ and $W(\lambda, x, x')$ depend only on geodesic distance $\sigma(x,x')$ due to homogeneity of $H_3$. This means we can set $x$ and $x'$ to correspond to the point $r=1$, where we normalized $\psi(r=1)$  in (\ref{normconv}).

On the other hand, an explicit expression for the heat kernel on $H_3$ was given in (2.9) of \cite{Giombi:2008vd}
\be K(t,\sigma) ={e^{-(M^2+1) t - {\sigma^2 \over 4t}} \over (4 \pi t)^{3/2}} {\sigma \over \sinh \sigma} \ , \ee
and so
\be W(\lambda,\sigma) =\int dt\, {e^{-(M^2+1) t + \lambda t- {\sigma^2 \over 4t}} \over (4 \pi t)^{3/2}} {\sigma \over \sinh \sigma} = {e^{-\sigma \sqrt{1 + M^2 - \lambda}} \over 4 \pi \sinh \sigma }\ , \ee
from which we can read off
\be  \mbox{Im} W(\lambda, \sigma=0) = \mbox{Im} {e^{-\sigma \sqrt{1 + M^2 - \lambda}} \over 4 \pi \sinh \sigma } \sim \mbox{Im}  \left({1 \over \sigma} - i\sqrt{\lambda - (1 + M^2)} + \ldots \right) = \sqrt{\lambda - (1 + M^2)} \ . \ee
The imaginary part of $W$ is seen as arising from the square root branch cut, and they are located precisely in the region  $\lambda > 1 + M^2$. 
We are also able to read off the density of state 
\be \rho(\lambda) = {1 \over \pi} \sqrt{\lambda - (1 + M^2)}~. \ee

The heat kernel is also useful for extracting the Green's function
\be G(x,x') = W(\lambda=0,\sigma(x,x')) = {e^{-\sigma \sqrt{1 + M^2}} \over 4 \pi \sinh \sigma } \ . \label{bulkbulk}\ee
One can easily check that
\be \Delta G(x,x') = {1 \over \sqrt{g}} \delta(x-x')~, \ee
and this Green's function is the same as (3.23) of \cite{Mann:1996ze}.

One can further check that the bulk-bulk Green's function (\ref{bulkbulk}) is consistent with the bulk-boundary Green's function typically used in the holographic contexts \cite{Witten:1998qj,Gubser:1998bc,Danielsson:1999zt,Hemming:2002kd} where one writes, (see e.g. (3.5) of  \cite{Hemming:2002kd})
\be G(y,w_+, w_-) \sim \left({y \over y^2 +\Delta w_+ \Delta w_-} \right)^{2 h_+}  \ , \label{bulkboundary}\ee
where $y$, $w_+$ and $w_-$ are Poincar\'{e} coordinates of $AdS_3$,
\be ds^2 = {dy^2 + dw_+ dw_- \over y^2} \ .\label{poincaremetric} \ee
To compare (\ref{bulkbulk}) and (\ref{bulkboundary}), we simply compute the geodesic distance between the points $(\epsilon,0,0)$ and $(y,w_+,w_-)$ with respect the metric (\ref{poincaremetric}). We find that
\be \sigma = \log\left({y^2 + w_+ w_- \over \epsilon y} \right) \ . \label{sigmayw}\ee
Substituting (\ref{sigmayw}) into (\ref{bulkbulk}) leads to (\ref{bulkboundary}) in the leading order in small $\epsilon$ expansion. 

Once these facts are established, it is straightforward to extend the result to the case of $H_3/Z$ by including image charges, as is often done. 

This confirms that harmonic analysis of continuum modes for $\lambda \ge 1 + M^2$ yields various observables that are consistent with various related quantities used in the literature. In a sense, our ``paradox'' can simply be considered the consequence of the false assumption that the harmonic analysis on $H_3/Z$ should be done in terms of discrete, normalizable modes. The fact that the eigenfunctions of $\Delta$ form a continuum after all was established long ago \cite{Camporesi:1991nw,Camporesi:1994ga}.\footnote{See also \cite{PerryWilliams,Kapec:2024zdj} for a more recent discussion.} It does mean, however,  that calculating the determinant as the product of eigenvalues is somewhat involved, since one must first regulate the spectrum to be countable and then take the continuum limit. The same can be said about the mode-by-mode computation of the partition function as a functional integral. But the consistency of the heat kernel and other formal methods \cite{Dunne:2007rt} is supported in part by obtaining consistent physical observables such as the Green's function. 

The story could have ended here, but we found an alternate, discrete, set of modes on $H_3$ and $H_3/Z$ for which quantities like the determinant and the Green's function can be computed yielding the same result. We will describe this alternate basis in the next section. 

\section{Surprise}\label{sec:surprise}

The reason we felt that the harmonic analysis on $H_3/Z$ should involve a discrete set of modes is because many work in the context of holography exploits the intuition that $AdS_3$ is like a box, with a discrete set of excitations. The poles in momentum space representation of the boundary-boundary two-point function such as (13) of \cite{Danielsson:1999zt} indeed exhibit such a structure. 

In this section, we will describe an alternate set of modes that appears to recover the determinant and the Green's function as the continuum modes discussed in (\ref{sec:res}). Of course, these alternate sets of modes can not be the eigenfunctions of the Klein-Gordon operator, as this is an unambiguous notion. We will be considering the eigenfunctions of a different operator. It will turn out that this operator is known as the ``weighted Laplacian'' operator in the warped compactification literature \cite{DeLuca:2024fbc}. In this section, we will describe how we re-discovered the weighted Laplacian, and how it reproduces the observables such as the thermal partition function and the Green's function.

\subsection{Determinant and the Green's function}

Our starting point is the realization that on-shell modes in $AdS_3$ involves not the spectrum of $\Delta$ through (\ref{eigen1}), but rather the equation of motion $\Delta \Phi = 0$, which one might just as well write $r^2 \Delta \Phi = 0$. If one were to write an eigenequation for the operator $\tilde \Delta = r^2 \Delta$, it would read
\be r^2 \Delta \Phi_\lambda(x_i) = \lambda \Phi_\lambda(x_i) \ . \label{eigen2} \ee
If we separate variables with the same ansatz (\ref{parameterize}) as in section \ref{sec:puzzle}, the equation for $g(r)$  becomes 
\be -r^2 \left(r^2-1\right)
   g''(r)+r\left(1-3
   r^2\right) g'(r)+
   \left(\frac{m_1^2}{\beta^2}+\frac{m_2^
   2 r^2}{r^2-1}\right)g(r) +r^2 M^2
   g(r) 
= \lambda g(r) ~. \label{waveeq3a} \ee
Note that in this form, the eigenvalue $\lambda$ and the quantum number $m_1^2 /\beta^2$ are additive, unlike in (\ref{waveeq3}).  The set of eigenvalues for $g(r)$ decaying at large $r$ and regular at $r=1$ is easy to find.  If one sets $\lambda = 0$ and $E = i m_1 / \beta$, then this is precisely the equation of motion for the spectrum of fluctuations on $AdS_3$ where the allowed values of $E$ was given in (\ref{Ellp}) as was reported, for instance, in \cite{Fitzpatrick:2010zm}. This means that the eigenvalues of $r^2 \Delta$ are given by 
\be \lambda = {m_1^2 \over \beta^2} + (2h + |m_2| + 2N)^2 ={m_1^2 \over \beta^2} + (2h + \ell + \ell')^2~. \ee
where
\be m_2= \ell - \ell'~, \qquad N = {1 \over 2} \left( \ell + \ell' - |\ell - \ell'| \right)~. \ee
Note that this quantity is manifestly real, and, the product over $m_1$, $\ell$, and $\ell'$ can be computed to yield (\ref{ZAH3}) which is in agreement with the thermal partition function on $AdS_3$. Specifically
\be ( \det r^2 \Delta)^{-1/2}= \prod_{m_1, \ell, \ell'}  \left({m_1^2 \over \beta^2} + (2h + \ell + \ell')^2\right)^{-1/2} = \prod_{\ell, \ell'} {1 \over 1 - q^{2h + \ell + \ell'} } 
\ . \label{ZAH3b} \ee
in agreement with (\ref{ZAH3}). 

It is useful to formalize the orthonormal properties of eigenmodes of (\ref{waveeq3a}). Let us set $m_1=0$ since it can be absorbed into $\lambda$. Let us label the discrete solutions by $N$. Since (\ref{waveeq3a}) also depends on discrete parameter $m_2$, let
\beq && -r^2 \left(r^2-1\right)
   g_{m_2,N}''(r)+r\left(1-3
   r^2\right) g_{m_2,N}'(r)+
   \left(\frac{m_2^
   2 r^2}{r^2-1}\right)g_{m_2,N}(r) +r^2 M^2
   g_{m_2,N}(r)\nonumber \\
&&   = \lambda_{m_2,N} g_{m_2,N}(r) ~. \label{waveeq3b} \eeq
We also have that 
\be g_{m_2,N}(r) = g_{-m_2,N} (r)~,\ee
since (\ref{waveeq3b}) is even in $m_2$. These eigenfunctions are orthogonal. The orthogonality relation is
\be \int dr \, {1 \over r} g_{n,N'}(r) g_{n,N}(r) = \delta_{N-N'} \ . \label{gortho1} \ee
The non-trivial element in this expression is the measure with respect to the integration by $r$. This can be determined by standard  Sturm-Liouville analysis. The same orthonomral basis was identified in \cite{Avis:1977yn,Burgess:1984ti}. An independent derivation of this fact is also presented in Appendix \ref{appa}.  We have also specified the normalization of $g_{m_2,N}$ by setting the coefficient of the Kroneker delta function in (\ref{gortho1}) to one. It then follows that
\be \sum_N g_{m_2,N}(r) g_{m_2,N}(r') = r \delta(r-r') \ , \ee
so that
\be \sum_N g_{m_2,N}(r') \int dr\, {1 \over r} g_{m_2,N}(r) g_{m_2,N'}(r) = g_{m_2,N'}(r')~, \ee
from either order of summing over $N$ and integrating over $r$. 

A natural Green's function to construct in terms of these objects is
\be G(x,x')  = \sum_{m_2,N} \int {dE \over 2 \pi} {1 \over  E^2 + (2 h + \ell + \ell') } e^{i E \tau+ i m_2 \theta} g_{m_2,N}(r)  e^{-i E \tau'- i m_2 \theta'} g_{m_2,N}(r') \ . \label{Green2}\ee
If one integrates out $E$, this is essentially (5.11) of \cite{Avis:1977yn} and (6) of \cite{Burgess:1984ti}.  The natural operator to act on this Green's function is $r^2 \Delta$. Upon its action, we find
\beq r^2 \Delta G(x,x')  &=& \sum_{m_2,N} \int {dE \over 2 \pi}  e^{i E \tau+ i m_2 \theta} g_{m_2,N}(r)  e^{-i E \tau'- i m_2 \theta'} g_{m_2,N}(r') \cr
& = & r\delta(\tau-\tau') \delta(\theta-\theta') \delta(r-r') \cr
& = & {r^2 \over \sqrt{g}} \delta(x-x') \ . \label{checkGreen} \eeq
We therefore see that (\ref{Green2}) is the Green's function of $\Delta$. In other words, it is the same quantity as (\ref{bulkbulk}).  Remarkably, the same quantity $G(x,x')$ can be expanded either in terms of eigenfunctions $\psi_\lambda(x)$ of $\Delta$ as in (\ref{bulkbulk}) which are delta function normalizable and forms a continuum, and the eigenfunctions $e^{i E \tau+ i m_2 \theta} g_{m_2,N}(r)$ of $r^2 \Delta$ which are strictly normalizable and forms a discrete set. 

This Green's function is for uncompactified $H_3$. It is straightforward to infer the Green's function on $H_3/Z$ by summing over images
\be G^{H_3/Z} (x,x') = \sum_{m_1} G^{H_3} (\tau, \theta, r; \tau' + m_1 \beta , \theta', r')~. \ee

Perhaps it is not very surprising since
\be  G_{r^2 \Delta}(x,x') =  {G_{\Delta}(x,x') \over r'^2}~, \ee
are closely related objects. If one knows one, it is easy to infer the other.

In a sense, this also explains why the temperature-dependent part of the determinant of $\Delta$ and $r^2 \Delta$ agreed. The Green's function and the free energy are related by
\be {1 \over 2}  \log \det \Delta = \beta F = -{1 \over 2} \tr \log G~. \label{trG} \ee
The difference for this quantity between  $G_{r^2 \Delta}(x,x')$ and $G_{\Delta}(x,x')$ is an additive term
\be -{1 \over 2} \int dx\, \sqrt{g}  \log r^2 \sim \beta \int dr \, r \log r ~,\ee
which is expected to be proportional to $\beta$ and divergent. But this is simply an additive constant to $F$. As such, it can be interpreted as an overall shift in the energy which is not physical. 

\subsection{Path integral}

Next, we will show that the scalar path integral can be decomposed in the modes of $r^2 \Delta$.\footnote{See \cite{Storchak:1991ab} for a related discussion.} Let us start with the action
\be S = \int d\tau\, d\phi\, d r \, \sqrt{g_{\tau \tau} g_{\phi\phi} g_{rr}} (g^{\tau\tau} (\partial_\tau \Phi)^2+g^{\theta\theta} (\partial_\theta \Phi)^2+g^{rr} (\partial_r \Phi)^2 )~. \label{action} \ee
Consider the following ansatz
\be \Phi(\tau,\theta,r) = \sum_{m_1,m_2,N} a_{m_1,m_2,N} e^{i m_1 \tau/\beta} e^{i m_2 \theta}  g_{m_2,N}(r)~, \label{ansatz} \ee
where we define $g_{m_2,N}(r)$ is as was defined previously in (\ref{waveeq3b}). 
We have already seen that the eigenvalues are
\be \lambda_{m_2,N} = (\sqrt{1 + M^2}+1 + |m_2| + 2 N)^2 = (2h + \ell + \ell')^2 \ . \ee

When the ansatz (\ref{ansatz}) is substituted into (\ref{action}), we obtain
\beq S &=&   \sum_{{m_1,m_2,N, \atop m2', N'}} \int dr \, d \theta \,\left[ {1 \over r} a^*_{m_1,m_2',N'}a_{m_1,m_2,N}{m_1^2 \over \beta^2} e^{-i m_2' \theta} g_{m_2',N'}(r) e^{i m_2 \theta} g_{m_2,N}(r) \right. \cr
&& \qquad\left. + r a^*_{m_1,m_2',N'}a_{m_1,m_2,N}  e^{-i m_2' \theta} g_{m_2',N'}(r) \Delta  e^{i m_2 \theta} g_{m_2,N}(r)\right]~, \eeq
where here, $\Delta$ only acts on the $(r,\theta)$ coordinates. With a slight re-writing, we have
\beq S &=&   \sum_{{m_1,m_2,N, \atop m2', N'}} \int dr \, d \theta \, {1 \over r} a^*_{m_1,m_2',N'}a_{m_1,m_2,N} \left({m_1^2 \over \beta^2} e^{-i m_2' \theta} g_{m_2',N'}(r) e^{i m_2 \theta} g_{m_2,N}(r) \right. \cr
&& \qquad \qquad\left. +  e^{-i m_2' \theta} g_{m_2',N'}(r) r^2 \Delta e^{i m_2 \theta} g_{m_2,N} \rule{0ex}{3ex}\right)~.\eeq
Now, using (\ref{waveeq3b}), we can write
\be S =  \sum_{{m_1,m_2,N, \atop m2', N'}} \int dr \, d \theta \, {1 \over r} a^*_{m_1,m_2',N'} a_{m_1,m_2,N} \left({m_1^2 \over \beta^2}+ \lambda_{m_2,N}\right) e^{-i m_2' \theta} g_{m_2',N'}(r) e^{i m_2 \theta} g_{m_2,N}(r)~. \ee
Since  $g_{m_2,N}(r)$ satisfies the orthogonality relation (\ref{gortho1}),  the action can be written
\be S =  \sum_{m_1,m_2,N}  a^*_{m_1,m_2,N} a_{m_1,m_2,N} \left({m_1^2 \over \beta^2}+ \lambda_{m_2,N}\right) ~,\label{actionfin} \ee
which then makes it manifest that
\be \det (r^2 \Delta) = \prod_{m_1, m_2, N} \left({m_1^2 \over \beta^2} + (2h + \ell + \ell')^2 \right) ~,\ee
and is in agreement with (\ref{ZAH3}) and (\ref{ZAH3b}).

Note that (\ref{actionfin}) makes manifest the interpretation of small fluctuations of scalar fields in $AdS_3$ as consisting of an infinite collection of simple harmonic oscillators. 

\subsection{Warped compactification and the weighted Laplacian} 

In this subsection, we will explain how the utility of operator $r^2 \Delta$ could have been anticipated based on previous work on warped compactifications. Warped compactification is a setup where one considers a space-time of the form 
\be ds^2 = e^{2A}(ds^2_{M_d}+ ds^2_{X_{D-d}}) \ , \ee
where $X$ is the internal manifold, and we are interested in the low energy dynamics from the perspective of $M_d$, but $A$ depends on the coordinates of $X$. The quantity $e^{2A}$ is generally referred to as the ``warp factor.'' Typically, in the context of model building, one is interested in inferring the spectrum of Kaluza-Klein-like particles arising from the normal modes on $X$, accounting for the complexity of the warp factor.

Precisely in this context, it was explained in  \cite{DeLuca:2024fbc} that the Kaluza-Klein spectrum is encoded not in the standard Laplacian, but instead the weighted Laplacian\footnote{We use the notation that $\hat \Delta_f$ also includes the mass term unlike in \cite{DeLuca:2024fbc}.}
\be  \hat \Delta_f \psi = (e^{-f} \hat \nabla^m e^f \hat \nabla_m + e^{2A} M^2)\psi~, \qquad f = (D-2) A~, \ee
where $\hat \nabla_m$ is the covariant derivative associated with the metric $ds^2_{X_{D-d}}$. We can see that (\ref{h3metric}) is precisely of this type when written in the form
\be ds_{H_3}^2 = r^2 \left(d \tau^2 + ds^2_{X_2}\right)~, \qquad ds_{X_2}^2 = {(r^2 - 1) \over r^2} d \theta^2 + {dr^2 \over r^2(r^2 - 1)} ~,\label{Gmetric}\ee
where $X_2$ has the topology of a disk, $d=1$, and $D=3$.   One can see that on the $(r,\theta)$ plane, 
\be r^2\Delta= e^{2A}\Delta =  \hat \Delta_f \ . \ee
This confirms that reading off the spectrum on $AdS_3$ from the eigenvalues of $r^2 \Delta$ is consistent with the general analysis of \cite{DeLuca:2024fbc}.

\section{Discussions}\label{sec:discussion}

In this article, we investigated the eigenvalues of Klein-Gordon operator $\Delta = -\nabla^2 + M^2$ on $H_3/Z$ with the goal of computing the determinant by computing their product. We approached this task with the prejudice that the set of eigenvalues must be discrete since one often thinks of $AdS_3$ as a box, and we expected that its Euclidean continuation with compactified time coordinate must still be like a box. We were surprised to find that normalizable eigenfunctions do not exist on $H_3/Z$ and from the careful analysis of the harmonic functions, we re-discovered that the eigenfunctions of Klein-Gordon operators form a continuum and are only normalizable in a delta-function sense. With these continuum sets of eigenfunctions, one can successfully compute the determinant and the Green's function both of which properly capture the spectrum of normalizable on-shell states on $AdS_3$.

Surprisingly, we found an alternative, discrete basis that also captures the determinant and the Green's function of the Klein-Gordon operator. The alternate basis are eigenfunctions of the operator $r^2 \Delta$. The determinant of this operator on $H_3/Z$ turns out to match the determinant computed in terms of the continuum modes in their dependence on the periodicity $\beta$ of the Euclideanized time coordinate.  We also found that the Green's function of $\Delta$ can be expressed in a form (\ref{Green2}) resembling spectral decomposition of $r^2 \Delta$. It was rather strange  for a Green's function of one operator is expressible as a spectral decomposition of another operator, but this appears to be a fact demonstrable by explicit computation. Most of these facts are known in fragmented form in various literature, but the holistic understanding of these relationships has not been made explicit to the best of our knowledge. It turns out that $r^2 \Delta$ is a special operator known as the ``weighted Laplacian'' in the context of warped compactification \cite{DeLuca:2024fbc}. 

One can understand the agreement of the determinant of $\Delta$ and $r^2 \Delta$ as a consequence of the product rule $\det (r^2 \Delta) = \det r^2 \det \Delta$ and the fact that $\det r^2$ has the effect of shifting the energies by a constant in the Boltzmann sum interpretation of the partition function. So the existence of this alternative is not so surprising after all. Once one understands how and why they agree, one is free to choose between the two by convenience. The spectroscopy of $\Delta$ and $r^2 \Delta$ have distinct advantages. The operator $\Delta$ preserves the homogeneity of the $H_3/Z$ geometry, but leads to a continuous spectrum. The operator $r^2 \Delta$, on the other hand, breaks the homogeneity but leads to a discrete spectrum. The determinant and the mode-by-mode evaluation of the partition function as a functional integral is easier to compute using $r^2 \Delta$. 

We have also examined the spectral decomposition of $\Delta$ in terms of normal modes of the BTZ geometry where one treats the coordinate $\theta$ as the Euclideanized time coordinate. This is equivalent to considering the eigenfunctions of $(r^2-1) \Delta$ operator. See appendix \ref{appb} for details. 

Our initial motivation was to try to make sense of the path integral of scalar fluctuations in the BTZ background. BTZ is an asymptotically $AdS_3$ space-time with a large black hole inside it. Understanding field fluctuations on this background is an important issue in quantum gravity. It has been known for a long time that making thermodynamic sense of these fluctuations is subtle since the spectrum of low energy oscillations in a black hole background generically forms a dense set \cite{tHooft:1984kcu}. See also e.g.\ \cite{Giveon:2019twx}. The reason for the dense spectrum is that waves oscillate indefinitely near the horizon. (Alternatively, one can show that there is a semi-infinite flat potential for the fluctuating fields in the tortoise coordinate where the horizon is at an infinite distance away.) Fluctuating fields are expected to contribute to the entropy of the black holes at order $\hbar^0$ where the leading contribution due to Bekenstein-Hawking formula is at order $\hbar^{-1}$. The dense spectrum appears to imply that the coefficient of $\hbar^0$ contribution is infinite. Various approaches have been pursued to regulate this divergence \cite{tHooft:1984kcu,Susskind:1994sm}.

The $AdS_3 \leftrightarrow H_3/Z \leftrightarrow BTZ$ correspondence provides an ``answer'' to this puzzle in that a finite expression is computed as the path integral of fields on $H_3/Z$. There remains a task of interpreting this quantity in the context of BTZ, but the most natural one is to take it literally as the $\hbar^0$ contribution to the free energy. In order to make this statement meaningful, it was important to relate the path integral on $H_3/Z$ to the determinant of $\Delta$. In light of what we explained in this article, one can work either with the determinant of $\Delta$ or $r^2 \Delta$. The latter makes the representation of degrees of freedom on $AdS_3$ as a collection of simple harmonic oscillators more manifest.

If one believes that we are correctly computing the path integral at order $\hbar^0$, it would be very interesting to explore other observables, such as the Hawking radiation \cite{Hartle:1976tp,Kraus:1994by,Kraus:1994fj,Vanzo:2011wq}, which is also believed to be a quantum effect.\footnote{Much of these earlier analysis appear to be carried out in the first quantized formalism. It should be possible to re-formulate these results in second quantized formalism using the results of this note.}

Authors of \cite{Denef:2009kn,Castro:2017mfj,Keeler:2018lza} successfully interpreted the inverse square root of the determinant of $\Delta$ as the generalization of the thermal partition function for a collection of damped simple harmonic oscillators. In appendix \ref{appb}, we show that the same quantity can be computed in terms of the continuous modes constituting the spectrum of $d/d\theta$. This equality is included in the right most column of figure \ref{figa}. It would be very interesting to understand the implications of this observation. These issues are explored in \cite{Law:2022zdq,Kapec:2024zdj}.

Finally, we note that the subtlety of warping we discussed in this note extends to $AdS_{d+1}$ and $AdS_{d+1}\mbox{-}BH$.  Unlike in the case of $d=2$, these two backgrounds are not related to each other via double Wick rotation, but they are separately related to a warped Euclidean geometry. Consider e.g. $EAdS_{d+1}$ whose metric we can write as
\be ds^2 = r^2 \left(d \tau^2  + {1 \over r^2(r^2 - 1)} dr^2+ {r^2 - 1 \over r^2} d \Omega_{d-1}^2 \right)\ . \ee
This is a geometry whose topology is that of $S_1 \times D_d$. The path integral of scalar fields in this background is once again the determinant of $r^2 \Delta$ in these coordinates.

For $AdS_{d+1}\mbox{-}BH$, its Euclidean continuation is
\be ds^2 = r^2 \left( d \Omega_{d-1}^2 + {dr^2 \over r^2 \left({r^2 \over b^2}+1 - {m \over r^{d-2}}\right)} + {1 \over r^2}\left({r^2 \over b^2} +1 - {m \over r^{d-2}}\right)d\tau^2 \right)\ . \ee
The topology of $(r,\tau)$ plane is that of a disk $D_2$. The topology of the Eucledian continuation is $S^{d-1} \times D_2$.  The ratio of the period of the radius of the disk to the radius of $S^{d-1}$ is given by \cite{Witten:1998zw}
\be \beta_0 = {4 \pi r b^2 r_+ \over n r_+^2+ (d-2) b^2}~, \qquad {r_+^2 \over b^2}+1 - {m \over r_+^{d-2}}=0 \ . \ee
The path integral of a scalar field in this background is again the determinant of $r^2 \Delta$ in these coordinates, and not that of $\Delta$. It is also interesting to note that the double analytic continuation will give rise to $dS_{d-1} \times D_2$ geometry.

Another interesting probe of the geometry is the spectral form factor \cite{Cotler:2016fpe} for the scalar field. Since spectral form factor complexifies the temperature $\beta \rightarrow \beta + i t$. Part of the issue in understanding the spectral form factor is to identify the contour in $(\beta,t)$ plane where the Hawking-Page transition is taking place. 

\section*{Acknowledgements} We are grateful to
N.~Berkovitz,
N.~Itzhaki, 
C.~Keeler, and especially 
S.~Giombi
for useful discussions. AH also thanks IFT-UNESP for hospitality where part of this work was done. The work of SC received funding under the Framework Program for Research and
“Horizon 2020” innovation under the Marie Sklodowska-Curie grant agreement $n^o$ 945298 and the Department of Physics at The Ohio State University. The work of SC  was also supported in part by the FACCTS Program at the University of Chicago.  The work of AH was supported in part by the U.S. Department of Energy, Office of Science, Office of High Energy Physics, under Award Number DE-SC0017647. The work of HN is supported in part by  CNPq grant 304583/2023-5
and FAPESP grant 2019/21281-4.
HN would also like to thank the ICTP-SAIFR for their support through FAPESP grant 2021/14335-0.

\section*{Appendix}

\appendix

\section{Schr\"{o}dinger form of the harmonic analysis \label{appa}}

\begin{figure}
\centerline{\includegraphics[width=3in]{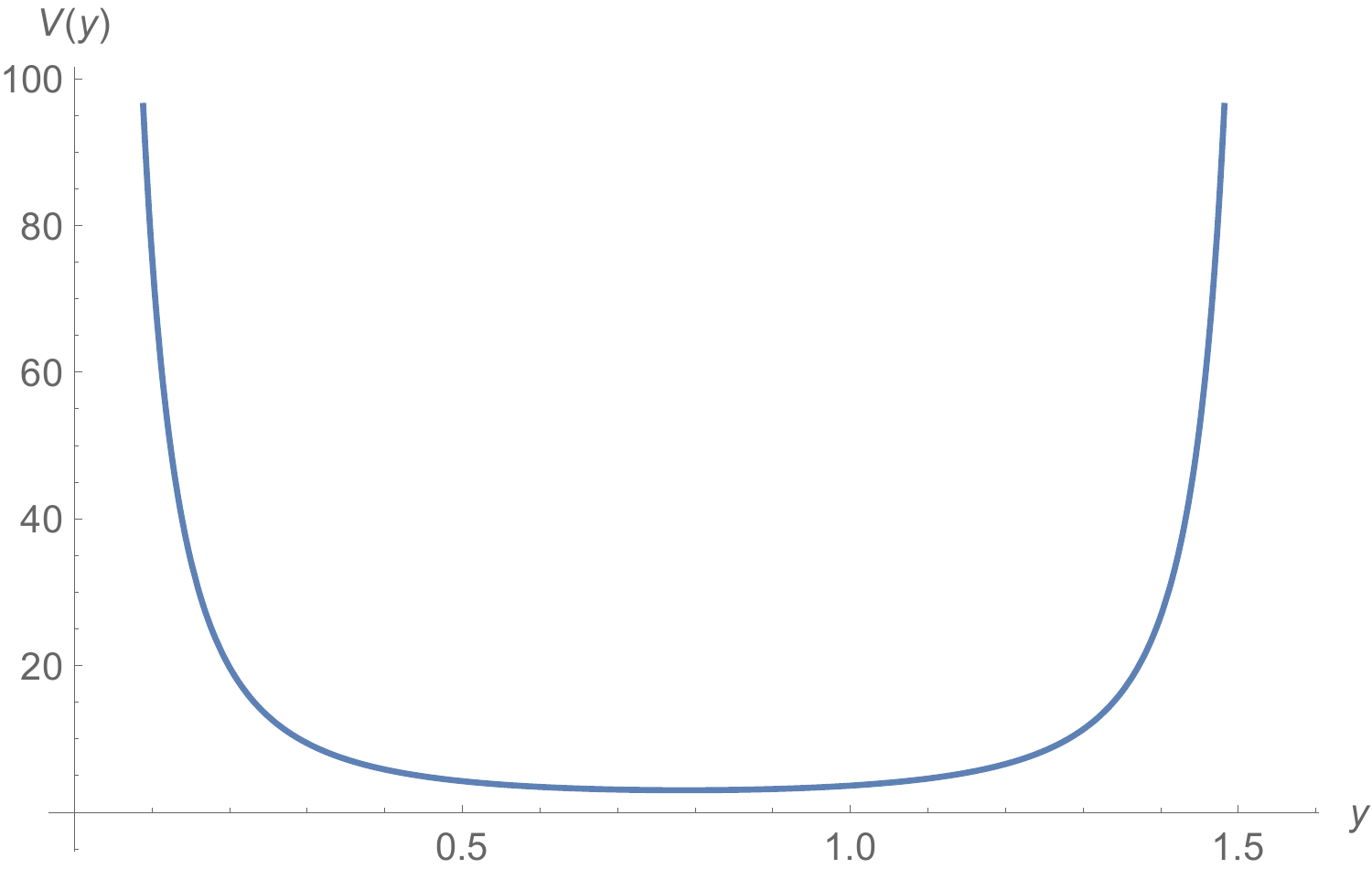}}
\caption{Potential $V(y)$ for $m_1=0$, $m_2=1$, $M^2=0$, and $\beta=1$. \label{fige}}
\end{figure}

In this appendix, we repeat the exercise of converting (\ref{waveeq3a}) into the Schr\"{o}dinger form so that we can read off the measure in $r$ for the orthogonality relation (\ref{gortho1}). We can set $m_1=0$ without loss of generality since it will simply shift $\lambda$.  This time, we use the coordinate
\be y =\arcsin\left({1 \over r} \right) \ ,\ee
and set
\be g(y) = f(y) \tilde g(y)~, \ee
where
\be f(y) = \sqrt{\tan(y)} = (r^2 - 1)^{-1/4}~. \ee
The equation for $\tilde g(y)$ then reads
\be - \tilde g''(y) + V(y) \tilde g(y) = \lambda \tilde g(y)~, \label{schrodinger} \ee
where
\be V(y) = \csc ^2(2 y) \left(-2 \left(m_2^2-1\right) \cos
   (2 y)+2 m_2^2+1\right) + M^2 \csc ^2(y) \ . \ee
This potential in the range $0 < y < \pi/2$, illustrated in figure \ref{fige},  is well-behaved and gives rise to a sensible bound state spectrum. 

Since $\tilde g(y)$ has eigenfunctions which is orthogonal with respect to measure $dy$, we expect $g(y)$ to be orthogonal with respect to the measure
\be {dy \over f(y)^2} ={dr \over r} \ . \label{measure}\ee
This confirms the measure of the orthogonality relation (\ref{gortho1}).  The same result can be obtained by performing the standard   Sturm-Liouville analysis.  

\section{Mode expansion and Green's function in BTZ background \label{appb}}

\begin{figure}
\centerline{\includegraphics[width=3in]{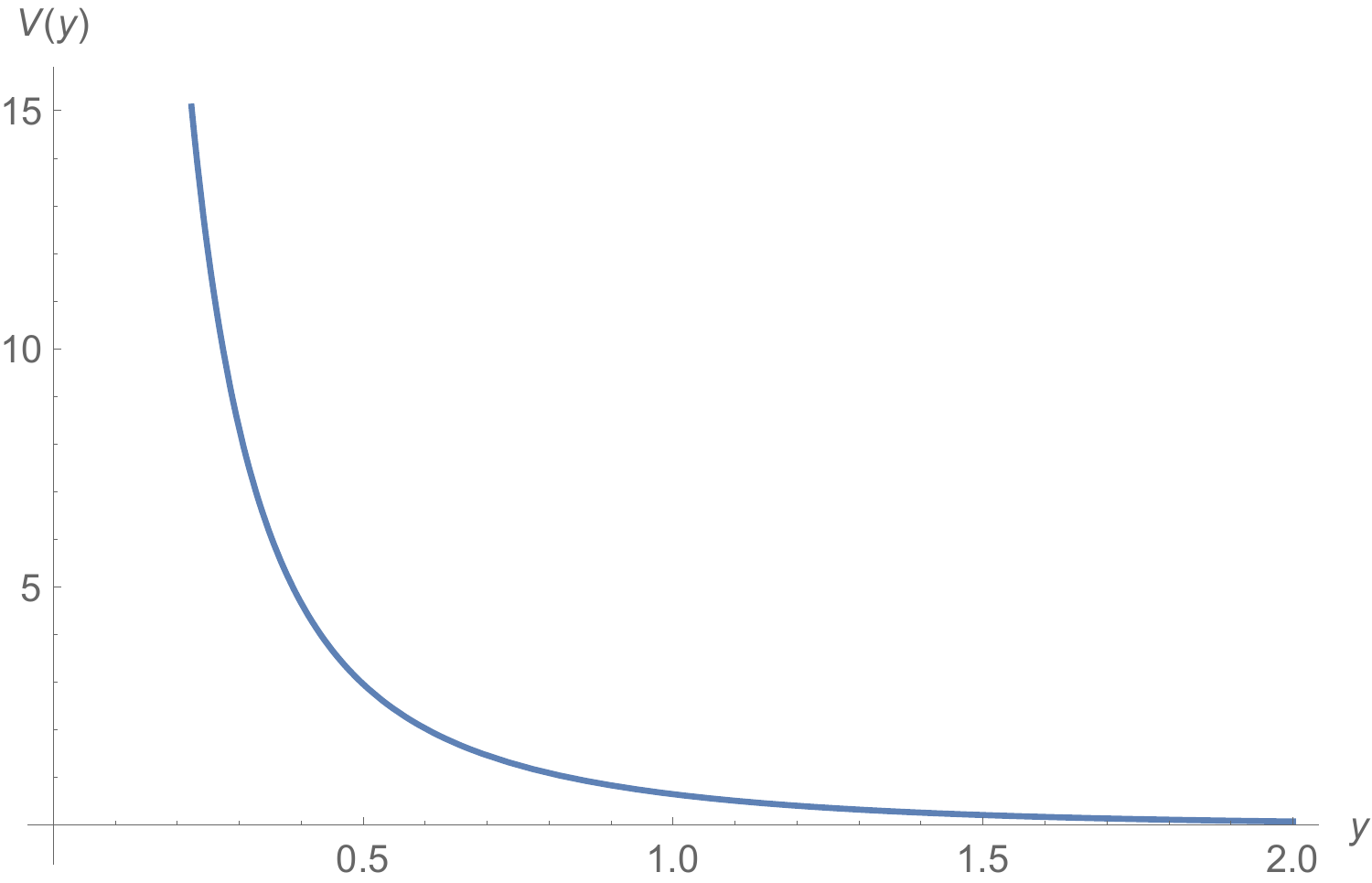}}
\caption{Potential $V(y)$ for $m_1=0$, $m_2=0$, $M^2=0$, and $\beta=1$. \label{figf}}
\end{figure}

In this appendix, we will describe the spectrum of fluctuation in BTZ background obtained by analytically continuing the $\theta$ coordinate. The starting point is the wave equation analogous to (\ref{waveeq3a}) corresponding to considering the spectrum of operator $\tsup{\Delta}=(r^2 - 1) \Delta$. 
\be - \left(r^2-1\right)^2
g''(r)+{r^2-1 \over r} \left(1-3  r^2\right) g'(r)+
\left(\frac{(r^2-1)m_1^2}{\beta^2 r^2}+m_2^2\right)g(r) + (r^2-1) M^2 g(r) 
= \lambda g(r) ~. \label{waveeq5a} \ee
where we absorb the $m_2^2$ into $\lambda$.  This equation is essentially obtained by multiplying the LHS of (\ref{waveeq3a}) by a factor of $(r^2 - 1)/r^2$. To write this equation in Schr\"{o}dinger form, we change variable to
\be y = {1 \over 2} \log\left({r-1 \over r+1}\right) \ . \ee
and
\be g(y) = f(y)  \tilde g(y)\ , \qquad f(y) = \sqrt{\tan y} \ . \ee
This leads to a Schr\"{o}dinger form (\ref{schrodinger}) with
\be V(y) = 
{m_1^2 \over \beta^2}\text{sech}^2(y) +(2 \cosh (2y)+1) \text{csch}^2(2 y) + M^2 \text{csch}^2(y)~,\ee
and has the form illustrated in figure \ref{figf}. Note that $y=0$ corresponds to $r=\infty$ and $y=\infty$ corresponds to $r=1$. The potential is such that we expect a continuous spectrum for $\lambda \ge 0$. This is the continuous spectrum `t Hooft regulated with a brick wall in \cite{tHooft:1984kcu}. 

We normalize the eigenmodes $g_{m_1, \lambda}(r)$ so that
\be \int dr\, {r \over r^2 - 1} g_{m_1, \lambda}(r) g_{m_1, \lambda'}(r)^* = \delta(\lambda- \lambda') \ . \ee
where the measure, like (\ref{measure}), is inferred from
\be {dy \over f(y)^2} ={r \over r^2 - 1} dr \ . \ee
From this, we infer that
\be \int d \lambda \, g_{m_1, \lambda}(r) g^*_{m_1, \lambda}(r')= {r^2 -1 \over r} \delta(r-r') \ . \ee

This leads one to consider the quantity
\be G(x,x')  = \sum_{m_1}\int d \lambda  \int {dE \over 2 \pi} {1 \over  E^2 + \lambda } e^{i m_1 \tau / \beta+ i E \theta} g_{m_1,\lambda}(r)  e^{-i m_1 \tau' / \beta- i E \theta'} g^*_{m_1,\lambda}(r') \ . \label{Green3}\ee
One can check that
\be (r^2 -1) \Delta G(x,x') = {r^2-1  \over \sqrt{g}} \delta(x-x')~, \ee
confirming that $G(x,x')$ is a Green's function of $\Delta$ just like (\ref{checkGreen}). The quantity (\ref{Green3}) is to be identified as the Green's function on the covering space of $H_3/Z$ where the $\tau$ coordinate is periodic with period $\beta$ but that the $\theta$ coordinate is decompactified. One must sum over images with $2 \pi$ periodicity in $\theta$ directions to compare with the Green's function on $H_3/Z$.

One can compute the determinant of $\Delta$ using the relation (\ref{trG}). First, note that
\be \log \det (r^2 - 1)\Delta= \log \det \Delta + \log \det (r^2 - 1) \ ,  \ee
and we discard the last term. We then see from (\ref{Green3}) that
\be -\log (\det(r^2 - 1) \Delta) =  \sum_{m_1}\int d \lambda \,  \rho_{m_1}(\lambda)\left(\log( 1 - e^{-2 \pi \lambda})+ \log(\lambda) \right) \ , \label{loglambda} \ee
where $\rho_{m_1}(\lambda)$ is defined by
\be \int dr\, {r \over r^2 - 1} g_{m_1, \lambda}(r) g_{m_1, \lambda}(r)^* =\rho_{m_1}(\lambda) \ ,  \ee
which includes an overall divergent volume factor. It also depends non-trivially on $\beta$. This is a quantity that is closely related to the spectral form factor. 
If one further rescales $\lambda =  \tilde \lambda / 2 \pi \beta$, we obtain
\be -\log (\det(r^2 - 1) \Delta) =  \sum_{m_1}\int {d \tilde \lambda \over 2 \pi \beta} \,  \rho_{m_1}(\tilde \lambda / 2 \pi \beta)  \left(\log(1 - e^{- \tilde \lambda/\beta})+ \log(\tilde \lambda/ 2 \pi \beta)   \right)\ , \ee
which is formally interpretable as the ``Boltzmann sum'' of continuum modes as was indicated in figure \ref{figa} except for the $\rho(\tilde \lambda /2 \pi \beta)\log (\tilde \lambda/2 \pi \beta)$ contribution.  In $AdS_3$, this extra term would have been temperature independent and would not have affected the thermodynamic observables. Here, $\rho(\lambda)$ is temperature dependent since the differential equation (\ref{waveeq5a}) depends expicitly on $\beta$. We suspect that there is a better way to understand these extra terms, but we will leave that consideration for future work. 

\providecommand{\href}[2]{#2}\begingroup\raggedright\endgroup

\end{document}